\documentclass{aastex63}
\usepackage{amsmath}
\usepackage{natbib}
\usepackage{multirow}
\usepackage{graphicx}
\usepackage{float}
\usepackage{amssymb}
\usepackage{threeparttable}
\usepackage{booktabs}
\usepackage{subfigure}

\hypersetup{linkcolor=magenta, citecolor=cyan, filecolor=yellow, urlcolor=blue}

\submitjournal{ApJ}

\shorttitle{Thermal component in Bright Short GRBs}
\shortauthors{Zhao et al.}

\begin{document}
\title{Detection of Prompt Fast-Variable Thermal Component in Multi-Pulse Short Gamma-Ray Burst 170206A}

\email{*qwtang@ncu.edu.cn; zouyc@hust.edu.cn; kaiwang@hust.edu.cn}

\author{Peng-Wei Zhao}
\affiliation{Department of Physics, Nanchang University, Nanchang 330031, P. R. China}
\author{Qing-Wen Tang}
\affiliation{Department of Physics, Nanchang University, Nanchang 330031, P. R. China}
\author{Yuan-Chuan Zou}
\affiliation{School of Physics, Huazhong University of Science and Technology, Wuhan 430074, P. R. China}
\author{Kai Wang}
\affiliation{School of Physics, Huazhong University of Science and Technology, Wuhan 430074, P. R. China}

\begin{abstract}

We report the detection of a strong thermal component in the short Gamma-Ray Burst 170206A with three intense pulses in its light curves, throughout which the fluxes of this thermal component exhibit fast temporal variability same as that of the accompanying non-thermal component. The values of the time-resolved low-energy photon index in the non-thermal component are between about -0.79 and -0.16, most of which are harder than -2/3 excepted in the synchrotron emission process. In addition, we found a common evolution between the thermal component and the non-thermal component, $E_{\rm p,CPL} \propto kT_{\rm BB}^{0.95\pm0.28}$, and $F_{\rm CPL} \propto F_{\rm BB}^{0.67\pm0.18}$, where  $E_{\rm p,CPL}$ and $F_{\rm CPL}$ are the peak photon energy and corresponding flux of the non-thermal component, and $ kT_{\rm BB}$ and $F_{BB}$ are the temperature and corresponding flux of the thermal component, respectively. Finally, we proposed that the photospheric thermal emission and the Comptonization of thermal photons may be responsible for the observational features of GRB 170206A.

\end{abstract}

\keywords{gamma-ray burst: general methods: data analysis: radiation mechanisms: thermal}

\section{Introduction \label{sec:Introduction}}

Gamma-ray bursts (GRBs) are believed to arise from the
deaths of massive stars or the coalescence of two compact stellar
objects such as neutron stars or black holes, which have both been
followed by an expanding fireball with a jet.
Many GRBs observed by several missions suggest
the prompt gamma-ray emission to be highly non-thermal~\citep{Mazets1981,Fenimore1982,Matz1985,Kaneko2006,Goldstein2012} and generated by synchrotron radiation of accelerated electrons in intense magnetic fields~\citep{Rees1994,Katz1994,Tavani1996,Sari1996,Sari1998}.
The values of the low-energy photon spectral index ($\alpha$), that is harder than -2/3 from
the observed GRB spectra, are different from the theoretical predictions. This low-energy photon index is expected to be -3/2 when the electrons undergo
the fast-cooling synchrotron, while it is about -2/3 when the
electron spectrum follows a slow-cooling synchrotron emission~\citep{Sari1998}.
A few theoretical models have been proposed to reconcile
the observed GRB prompt spectra with the synchrotron process.
Some of them invoke effects that produce a hardening of the
low-energy spectral index, such as a decaying magnetic field
~\citep{Peer2006a,Uhm2014,Zhang2020,Wang2021}, inverse Compton
scattering in the Klein--Nishina regime or a marginally fast cooling regime~\citep{Derishev2001,Nakar2009,Wang2009,Daigne2011}.

Actually, the emission from this fireball is expected to be thermal, which is
originated from the non-dissipative photosphere~\citep{Goodman1986,Paczynski1986,Rees1994,Ryde2004,Peer2008,Beloborodov2010,Peer2011,Beloborodov2011,Ghirlanda2013,Larsson2015,Ryde2017}.
This pure thermal component fitted by
a standard Planck blackbody function (hereafter BB) is found
in many Fermi-GBM GRBs, such as GRB 150101B and other GRBs~\citep{Burns2018,Acuner2019,Acuner2020}.
Even the low-energy photon index is acceptable in the synchrotron theory, the modified thermal
processes are proposed to account for the observations, such as the dissipative
photosphere~\citep{Rees2005,Veres2012,Giannios2012,Lundman2013,Lundman2014,Lundman2018}.
A trend has therefore evolved with the possibility of reconciling
synchrotron emission with the $\alpha$ distributions, which
consists of fitting a blackbody (non-dissipative photosphere) in combination
with the typically fitted non-thermal spectral function to spectra observed
by Fermi/GBM~\citep{Ryde2005,Battelino2007,Guiriec2011,Axelsson2012,Guiriec2013,Iyyani2013,Preece2014,Burgess2015,Tang2021}.

GRB 120323A was the first short GRB (SGRB) with contemporaneous detection of the thermal component and non-thermal component in the prompt phase, with the single-pulse lightcurves and a decaying pattern both in its thermal flux and thermal temperature~\citep{Guiriec2013}. Among the top ten brightest fluence-selected SGRBs detected by Fermi/GBM as of 2021 December, we have searched for such spectral property, and found that GRB 170206A with a strong thermal-component detection, which however shows very different properties from GRB 120303A, such as the tracking pattern between the thermal component and non-thermal component. In this work, we report the results of this unique SGRB and explore its possible physical origins.
This paper is organized as follows.
In \S 2, we present the observations of GRB 170206A.
In \S 3, data analysis of GRB 170206A and the results are presented.
In \S 4, we discuss the origins of these two spectral components.
The conclusion and discussion are presented in \S 5.

\section{Observations \label{sec:Observation}}
GRB 170206A was triggered at 10:51:57.70 UT on 06 February 2017 ($T_0$) by Fermi Gamma-Ray Burst Monitor (GBM) with R.A.$_{\rm GBM}= 211.80^{\circ}$, Decl.$_{\rm GBM}= 13.06^{\circ}$ and 1 sigma uncertainty of 1.14 degrees. The GBM light curve shows a short, bright burst with a duration of about 1.2 s in the energy range of 50-300 keV~\citep{GBMGCN}.
It was also detected by Fermi Large Area Telescope (LAT) with the best location at R.A.$_{\rm LAT}= 212.79^{\circ}$, decl.$_{\rm LAT}= 14.48^{\circ}$ and the 90\% containment statistical error radius 0.85 degrees, which is consistent with the GBM position. The angle from the Fermi/LAT boresight at the GBM trigger time ($T_0$) is about $67^{\circ}$, the highest-energy photon detected by LAT is about 811 MeV event which is observed 3.17 seconds after the GBM trigger~\citep{LATGCN}.

GRB 170206A was detected by Konus-Wind, INTEGRAL/SPI-ACS, and Mars-Odyssey/HEND, with the center location of R.A.$_{\rm IPN}$ = 212.63$^{\circ}$ and decl.$_{\rm IPN}$ = 14.24$^{\circ}$~\citep{KonusWindGCN,IPNGCN}. POLAR on-board the Chinese space laboratory Tiangong-2 detected it in the energy range of about 20-500 keV, which shows that GRB 170206A consists of multiple peaks and with the minimum detectable polarization of about 5.7\% ~\citep{POLARGCN}.

\section{Data Analysis \label{sec:Data Analysis}}

\subsection{Event selections}

For the GBM data, three NaI detectors most close to the GRB position ($n9, na$ and $nb$) and one BGO detector ($b1$) with the lowest angle of incidence are included. For the time-tagged event (TTE) from these NaI detectors employed in the following sections, we ignore the last two channels and events with photon energy less than 8 keV. For TTE data of the BGO detector, the channels with energy below 200 keV and above 40 MeV are ignored. We choose the time intervals of [-25 s, -10 s] and [15 s, 30 s] away from the GBM trigger time to fit the background. Instrument response files are selected with the \textit{rsp2} files throughout the data analysis.

For the LAT data, the LAT--\textit{Transient020E} events with a zenith angle cut of 100$^{\circ}$ are selected, whose energy are between 100~MeV to 10~GeV. Region of interest (ROI) is chosen within the radius of 12$^{\circ}$ from the Fermi/LAT localization, such as R.A.$_{\rm LAT}= 212.79^{\circ}$, decl.$_{\rm LAT}= 14.48^{\circ}$.

\subsection{Temporal analysis \label{sec:timing}}

\begin{figure*}
\centering
\includegraphics[width=0.70\textwidth]{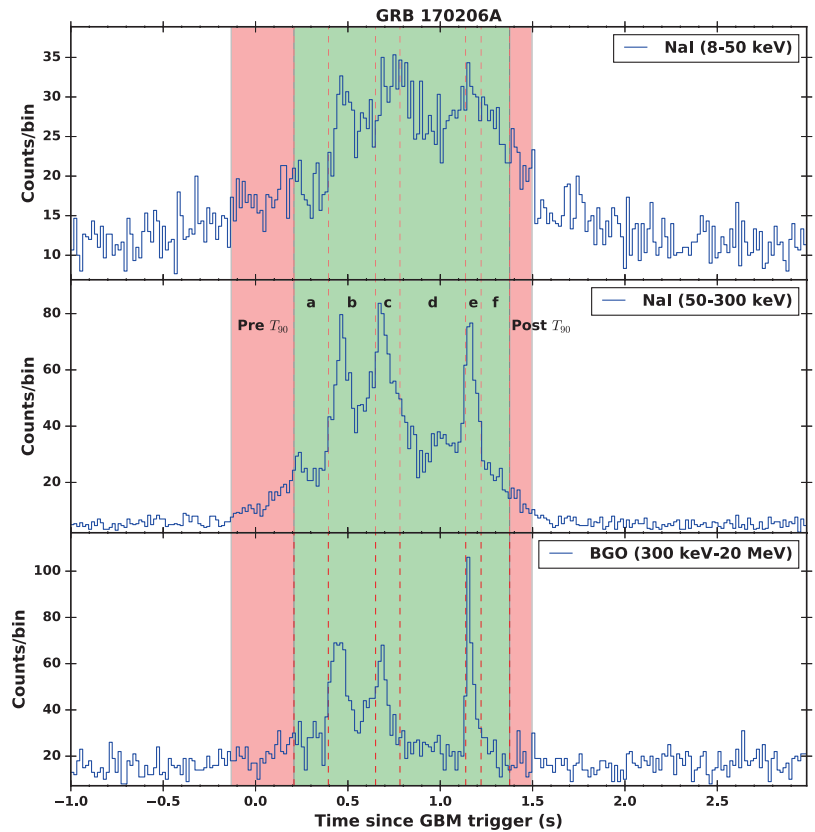}
\caption{Composite light curves for GRB 170206A. From top to bottom, low energy band lightcurve (8-50 keV, LE band), GBM $T_{90}$ band lightcurve (50-300 keV, $T_{90}$ band), the main BGO energy band lightcurve (BGO band), and the LAT lightcurve (100 MeV--10 GeV, LAT band). The green shadowed region covers the GBM $T_{90}$ period, the red shadows before and after which are the Pre-$T_{90}$ period, Post-$T_{90}$ period respectively, detail please see the text in Section \ref{sec:timing}. The red dashed vertical lines divide the GBM $T_{90}$ into six time-resolved epochs, which are labeled from a to f.}
\label{fig:lc1}
\end{figure*}

We built the multi-wavelength GBM light curves as well as the LAT light curves, which are shown in Figure \ref{fig:lc1}.

For the GBM light curves, we plotted them in three energy bands, the low energy band (8-50 keV, hereafter LE band), the energy band employed to estimate the GBM $T_{90}$ (50-300 keV, hereafter $T_{90}$ band), among which 90\% of the burst's fluence was accumulated, and the main energy range of the BGO detector (300 keV-20 MeV, hereafter BGO band). For those light curves in LE band and $T_{90}$ band, the average count rates of three NaI detectors ($n9, na$ and $nb$) are calculated.
As seen in Figure \ref{fig:lc1}, light curves both in $T_{90}$ band and BGO band show fast-variable property with three intensive pulses, while light curve in LE band can be also distinguished by three pulses. In order to perform the time-resolved spectral analysis in the following sections, six epochs are finally derived by rebinning the TTE data of the brightest NaI detector ($n9$) using the Bayesian Blocks method (BBlocks; \citealt{Scargle2013}) with a false alarm probability of $p_0 = 0.001$, which is the chance probability of the correct bin configuration. The derived time-resolved epochs are plotted with the red dashed vertical lines and labeled from epoch a to epoch f, amongst which the epochs b, c, and e are dominated by the first pulse (P1), the second pulse (P2), and the third pulse (P3) as seen in Figure \ref{fig:lc1}.

In order to discuss the spectral properties before and after the GBM $T_{90}$ period (epochs a, b, c, d, e and f), we perform the same BBlocks analysis as above in the time intervals [-0.500 s, 0.208 s], [1.376 s, 2.000 s] relative to $T_0$. As a result, we derived two periods nearest the $T_{90}$, such as Pre-$T_{90}$ period of [$T_{0}$-0.133, $T_{0}$+0.208] and Post-$T_{90}$ period of [$T_{0}$+1.376, $T_{0}$+1.497], which are also employed to perform time-integrated spectral analysis in the following sections.

As for the LAT data, we perform the unbinned likelihood analysis in the time range of 1 second before and 100 seconds after the GBM trigger time, and calculate the probability of each photon associated with GRB 170206A by $Fermi$ Science Tools ($gtsrcprob$). As seen in Table \ref{tab:photon}, there are six high energy photon events detected by Fermi/LAT, however, the only one photon within GBM $T_{90}$ has a probability less than 50\%, thus we did not include the LAT data in the following spectral analysis~\citep{LATGCN,Ackermann2013,Ajello2019}.

\begin{deluxetable*}{ccc}
	\tablewidth{0pt}
	\tablecaption{Properties of the high-energy photons of GRB 170206A detected by Fermi-LAT \label{tab:photon}}
	\tablehead{
		\colhead{Arrival Time\tablenotemark{a}}
		&\colhead{Photon Energy}
		&\colhead{Probability\tablenotemark{b}}\\
		\colhead{s}
		&\colhead{MeV}
		&\colhead{}
	}
	\startdata
	 0.85 & 121.7 & 36.97\% \\
	 3.17 & 810.6 & 99.99\% \\
	 6.60 & 389.0 & 96.27\% \\
	 61.33 & 306.3 & 65.62\% \\
	 82.51 & 105.9 & 5.55\% \\
	 98.25 & 121.5 & 18.97\% \\
	\hline
	\enddata
	\tablenotetext{a}{Arrival time of each high-energy photon after GBM $T_0$}
	\tablenotetext{b}{Probability of each high-energy photon that associated with GRB 170206A.}
\end{deluxetable*}

\subsection{Spectral analysis \label{specana}}
\subsubsection{General method}
Four models are defined to fit the gamma-ray data of GRB 170206A, namely, the cutoff power-law model (CPL), the Band model (BAND), the CPL+BB model and the BAND+BB model. For the latter two BB-joint models, the CPL+BB model consists of the CPL component and BB component while the BAND+BB model comprises the BAND component and the BB component. These models are expressed below:

(i) The BAND model, which is written same as that in \citep{1993ApJ...413..281B},
\begin{eqnarray}
N(E)_{\rm BAND} = A_{\rm BAND} \left\{ \begin{array}{ll}
(\frac{E}{100\ {\rm keV}})^{\alpha} e^{[-E(2+\alpha)/E_{p}]}, & E\leq \frac{\alpha-\beta}{2+\alpha}E_{p} \\
\\
(\frac{(\alpha-\beta)E_{p}}{(2+\alpha) 100\ {\rm keV}})^{(\alpha-\beta)} e^{(\beta-\alpha)}(\frac{E}{100\ {\rm keV}})^{\beta}, & E\geq \frac{\alpha-\beta}{2+\alpha}E_{p} \\
\end{array} \right.
\label{Eq:BAND}
\end{eqnarray}
where $\alpha$, $\beta$ are the low-energy photon index and the high-energy photon index respectively, and $E_{\rm p}$ (or $E_{\rm p, BAND}$) is the peak energy in the $\nu F_\nu$ spectrum.

(ii) The CPL model is written as
\begin{equation}
N(E)_{\rm CPL} = A_{\rm CPL} (\frac{E}{100\ {\rm keV}})^{\alpha} e^{-E/E_{\rm c}},
\label{Eq:CPL}
\end{equation}
where $\alpha$ is the photon index and $E_{\rm c}$ is the cutoff energy. The peak energy of the CPL model ($E_{\rm p, CPL}$) is calculated by $E_{\rm p, CPL}$ = $(2+\alpha)\times E_{\rm c}$.

(iii) The BB model is given by
\begin{equation}
	N(E)_{\rm BB} = A_{\rm BB} \frac{E^{2}}{{\rm exp}(E/kT_{\rm BB})-1},
	\label{Eq:BB}
\end{equation}
where $k$ is the Boltzmann's constant, and the joint parameter $kT_{\rm BB}$ as a output parameter in common. For above all models, $A$ is the amplitude. The free parameters in a candidate model are initialed at the typical spectral parameter values from the Fermi-GBM catalog~\citep{2020ApJ...893...46V} and allowed in the broad ranges.

Other six models are also included to make comparisons, such as the main models (BAND, CPL) with an additional power-law decay model (PL) or the multicolor blackbody (mBB), which are presented in Appendix~\ref{appendix}. As discussed in Appendix~\ref{appendix}, the most possible model mBB can not fit the SED well in $T_{90}$ period and three time-resolved spectra, such as epochs b,d and e, thus we did not present it in the following sections.

As a common method in the GBM spectral analysis, we employ the maximum likelihood estimate (MLE) method, which is suitable for the Poisson data and the Gaussian background (PG$_{\rm stat}$; \citealt{Cash1979}). For each fitting, a likelihood value $L(\vec{\theta})$ as the function of the free parameters $\vec{\theta}$ is derived, then the value of the Akaike Information Criterion (AIC; \citealt{1974ITAC...19..716A}), defined as AIC=-2ln$L(\vec{\theta})$+2$k$, and the value of the Bayesian Information Criterion (BIC; \citealt{1978AnSta...6..461S}), defined as BIC=-2ln$L(\vec{\theta})$+$m \ln n$, are calculated, where $m$ is the number of free parameters to be estimated and $n$ is the number of observations (the sum of the selected GBM energy channels). In this work, the Multi-Mission Maximum Likelihood package (3ML; \citealt{2015arXiv150708343V}) is employed to carry out all the spectral analysis and the parameter estimation.

In this paper, given any two estimated models, the preferred model is the one that provides the minimum BIC score. We use $\Delta$BIC to describe the evidence against a candidate model as the best model in the spectral analysis of GRB 170206A. With respect to the best model with the minimum BIC (BIC$_{\rm minimum}$), the evidence that the best model is against the candidate model is very strong when $\Delta$BIC (= BIC$_{\rm candidate}$ - BIC$_{\rm minimum}) > 10$  while $\Delta$BIC $> 6$ is strong~\citep{Kass1995}. Finally, if $\Delta$BIC is smaller than 6, the candidate model is classified as the compared model.

\subsubsection{Time-integrated spectral analysis}
We perform the time-integrated spectral analysis of GRB 170206A in three main time intervals, that is Pre-$T_{90}$ period, $T_{90}$ period and Post-$T_{90}$ period described in Section \ref{sec:timing}, whose results are presented in Table~\ref{tab:GRB170206A_spectra}.

\begin{figure*}
\centering
\includegraphics[width=0.45\textwidth]{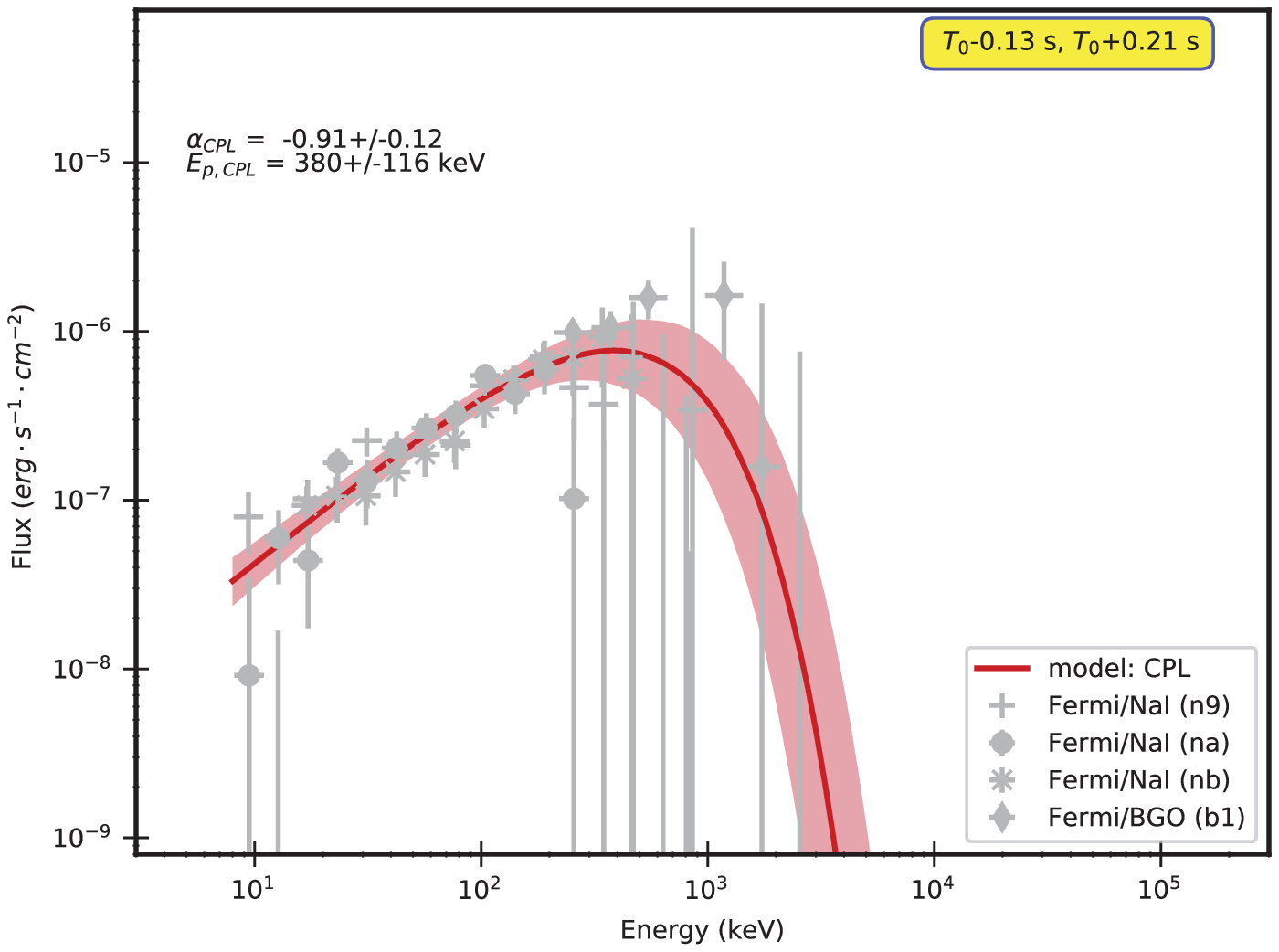}
\includegraphics[width=0.45\textwidth]{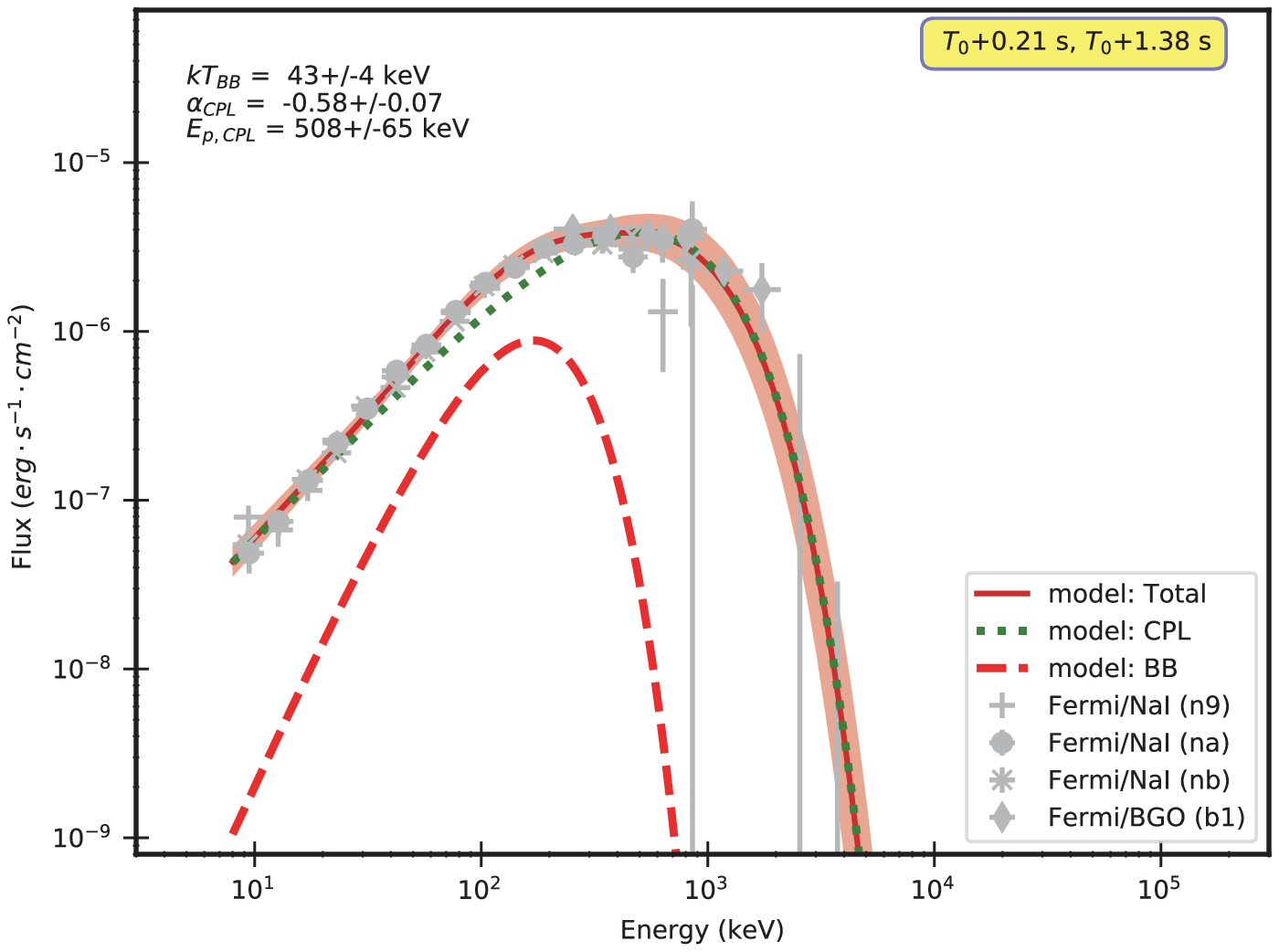}
\includegraphics[width=0.45\textwidth]{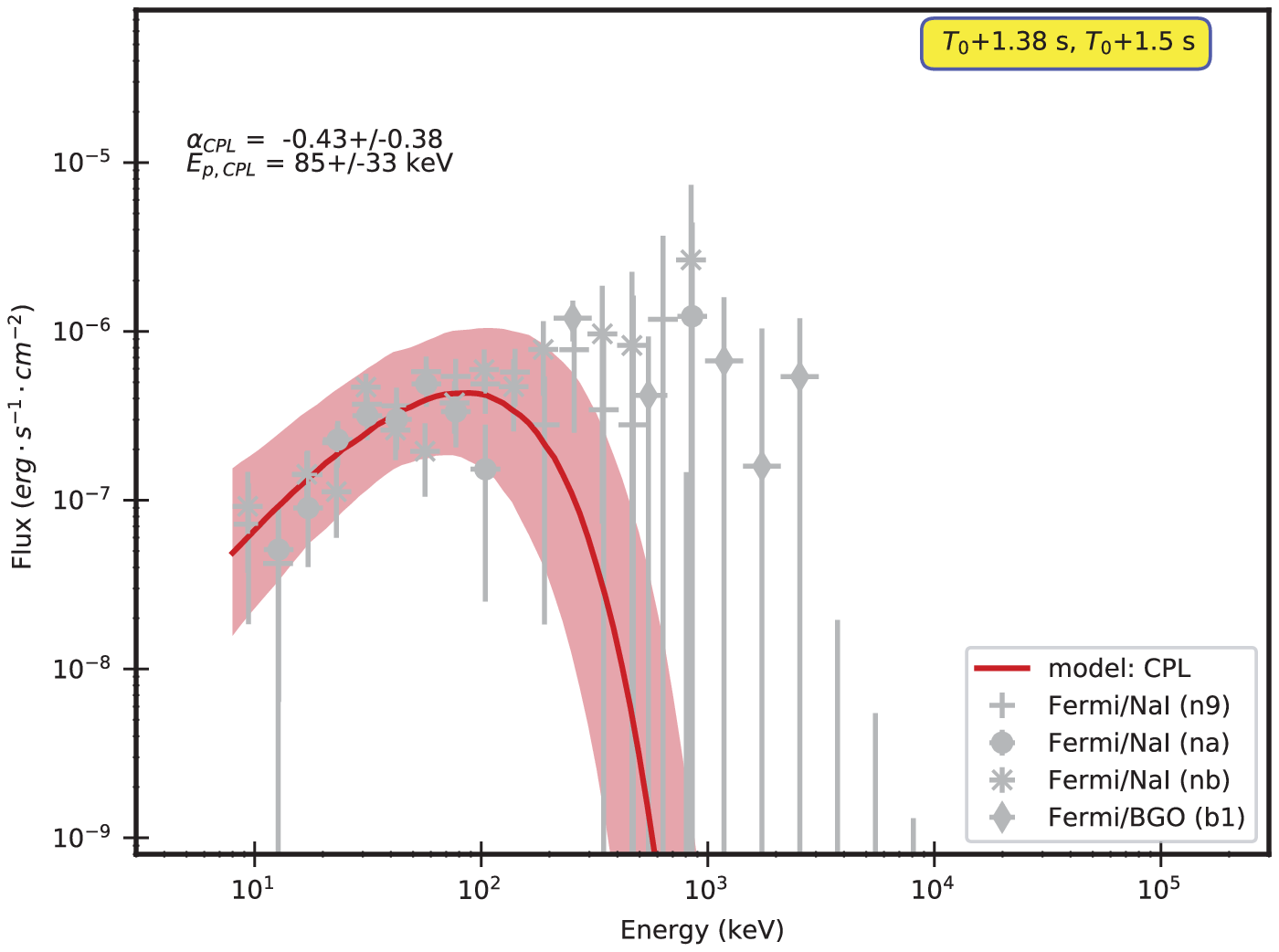}
\caption{Spectral energy distributions and best-fitted model for the time-integrated spectra of GRB 170206A. Top left: Pre-$T_{90}$ period between $T_0$-0.13 s and $T_0$+0.21 s.  Top right: $T_{90}$ period between $T_0$+0.21 s and $T_0$+1.38 s. Bottom: Post-$T_{90}$ period between $T_0$+1.38 s and $T_0$+1.50 s.
Data points are from the Fermi/GBM. For spectra best-fitted by the CPL model, the red solid line represents the resultant CPL model. For spectra best-fitted by the CPL+BB model, the green dotted line represents the CPL component, the red dashed line represents the BB component and the red solid line is the total modeled flux. All red shadow regions are the 95\% confidence intervals of the total modeled flux.}
\label{fig:sed1}
\end{figure*}

For the $T_{90}$ period, the BAND+BB model is not suitable to fit the gamma-ray data with an unconstrained $\beta$, e.g., $\beta < -5.0$. Note that, the BIC value in the CPL+BB model is also smaller by 6.1 than that in the BAND+BB model. The CPL+BB model has a $\Delta$BIC larger than other two models by 6, such as 7.5 with respect to the BAND model and 14.4 to the CPL model, thus is considered as the best-fit model. The energy fluxes of the CPL component and the BB component in the CPL+BB model are calculated in the energy range between 8 keV and 40 MeV, such as $F_{\rm CPL}$, $F_{\rm BB}$ of $(8.7\pm1.8)$ $\times {\rm 10^{-6} erg\ cm^{-2} \ s^{-1}}$, $(1.2\pm0.5)$ $\times {\rm 10^{-6} erg\ cm^{-2} \ s^{-1}}$ respectively. The BB component has about 12\% of the total modeled energy flux. The $\nu F_\nu$ peak energy of the CPL component is 508$\pm$65 keV while the BB component has a temperature of $kT_{\rm BB} =$ 43$\pm$4 keV.
The $\nu F_\nu$ spectral energy distribution (SED) fitted by the CPL+BB model is plotted at the top right of Figure~\ref{fig:sed1}.

For the Pre-$T_{90}$ period, one can see that the CPL model is suited for fitting the gamma-ray spectrum with the $\Delta$BIC larger than 6 compared to the other three models, thus the CPL model is the best-fit model in the Pre-$T_{90}$ period, which is plotted at the top left of Figure~\ref{fig:sed1}.

For the Post-$T_{90}$ period, the parameters could not be constrained well in both BAND and BAND+BB models. The CPL model is the better model to fit the data comparing with the CPL+BB model, e.g., $\Delta$BIC = 6.1. Therefore, the best-fit model for the Post-$T_{90}$ period is the CPL model, which is plotted at the bottom of Figure~\ref{fig:sed1}.

\begin{table*}
\caption{\label{tab:GRB170206A_spectra}Spectral-fitting results of GRB 170206A}
\begin{center}
{\tiny
\begin{tabular}{ccccccccc}
\hline
\multicolumn{1}{c}{Models} &\multicolumn{4}{c}{Main component} & \multicolumn{2}{c}{BB component} & \multicolumn{2}{c}{Stat. \& dof} \\
     &\multicolumn{4}{c}{BAND or CPL} & \multicolumn{2}{c}{ BB}  & \multicolumn{2}{c}{}\\
\cmidrule(lr){2-5} \cmidrule(lr){6-7} \cmidrule(lr){8-9}\\
\multicolumn{1}{c}{\textbf{T$_{\rm Start}$\ --\ T$_{\rm End}$}}    & \multicolumn{1}{c}{$E_{\rm p, main}$} & \multicolumn{1}{c}{$\alpha$} & \multicolumn{1}{c}{$\beta$} & \multicolumn{1}{c}{$F_{\rm main}$} & \multicolumn{1}{c}{$kT_{\rm BB}$} & \multicolumn{1}{c}{$F_{\rm BB}$} & \multicolumn{1}{c}{AIC/BIC/-log(likelihood)} & \multicolumn{1}{c}{dof}  \\
\multicolumn{1}{c}{\textbf{s \ \ -- \ \ \  s}}     & \multicolumn{1}{c}{keV} & \multicolumn{1}{c}{} & \multicolumn{1}{c}{} & \multicolumn{1}{c}{${\rm 10^{-6} erg\ cm^{-2} \ s^{-1}}$} & \multicolumn{1}{c}{keV} & \multicolumn{1}{c}{${\rm 10^{-6} erg\ cm^{-2} \ s^{-1}}$} & \multicolumn{1}{c}{} & \multicolumn{1}{c}{}  \\
\hline
$\textbf{Time-integrated}$&&&&&&\\	
$\textbf{Pre-$T_{90}$}$&&&&&&\\	
$\textbf{ -0.133 -- 0.208 }$&&&&&&\\	
BAND	&$	382	\pm	79	$ &$	-0.91	\pm	0.11	$ &$<-5$ &$	2.0	\pm	1.7	$ &-- &--&	 818.3	/	834.9	/	 405.1	&	 474	\\
BAND+BB	&$	10	\pm	10	$ &$	-1.02	\pm	0.21	$ &$-1.66	\pm	0.08$ &$	0.7	\pm	0.4	$ &$	58	\pm	7	 $ &$	 2.1	\pm	0.1	0.5	$&	836.7	/	861.6	/	412.3	&	472	\\
CPL	&$	380\pm	116	$ &$	-0.91	\pm	0.12	$ &-- &$	1.9	\pm	1.3$ &-- &--&	816.3	/	828.8	/	 405.1	 &	 475	\\
CPL+BB	&$	328	\pm	136	$ &$	-0.35	\pm	0.46	$ &-- &$	1.6	\pm	0.7	$ &$	9	\pm	2	$ &$<0.01 $&	 816.8	/	 837.6	/	403.4	&	473	\\
$\textbf{$T_{90}$}$&&&&&&\\	
$\textbf{ 0.208 -- 1.376}$&&&&&&\\	
BAND	&$	344	\pm	14	$ &$	-0.31	\pm	0.04	$ &$	-2.86	\pm	0.17	$ &$	10.8	\pm	1.0	$ &-- &--&	 3038.9	/	3055.5	/	1515.4	&	474	\\
BAND+BB	&$	508	\pm	41	$ &$	-0.58	\pm	0.06	$ &$<-5$ &$	8.9	\pm	1.4 $ &$	43	\pm	4	$ &$	1.2	 \pm	0.4	$&	 3029.1	/	3054.1	/	1508.6	&	472	\\
CPL	&$	379	\pm	17	$ &$	-0.39	\pm	0.03	$ &-- &$	9.1	\pm	0.8	$ &-- &--&	3049.9	/	3062.4	/	 1521.9	 &	 475	\\
CPL+BB	&$	508	\pm	65	$ &$	-0.58	\pm	0.07	$ &-- &$	8.7	\pm	1.8	$ &$	43	\pm	4	$ &$	1.2	 \pm	0.5	$&	 3027.1	/	3048.0	/	1508.6	&	473	\\
$\textbf{Post-$T_{90}$}$&&&&&&\\
$\textbf{ 1.376 -- 1.497}$&&&&&&\\
BAND	&-- &-- &-- &-- &-- &--&	 Unconstrained	&	474	\\
BAND+BB	&-- &-- &-- &-- &-- &--&	 Unconstrained	&	472	\\
CPL	&$	85	\pm	33	$ &$	-0.43	\pm	0.38	$ &-- &$	0.9	\pm	0.4	$ &-- &--&	-624.7	/	-612.2	/	 -315.4	 &	 475	\\
CPL+BB	&$	168	\pm	277	$ &$	0.27	\pm	3.35	$ &-- &$	2.7	\pm	0.01	$ &$	11	\pm	2	$ &$	0.4	 \pm	 0.1$&	-626.8	/	-606.1	/	-318.5	&	473	\\
\hline
$\textbf{Time-resolved}$&&&&&&\\	
$\textbf{(a) 0.208  -- 0.394}$&&&&&&\\	
BAND	&$	345	\pm	30	$ &$	-0.16	\pm	0.13	$ &$	-5.80	\pm	2.74	$ &$	4.7	\pm	1.0$ &-- &--&	 190.6	/	 207.3	/	91.3	&	474	\\
BAND+BB	&$	359	\pm	29	$ &$	2.29	\pm	1.85	$ &$<-5$ &$	3.9	\pm	2.2$ &$	24	\pm	5	$ &$	0.6	 \pm	0.4$&	 191.9	/	216.9	/	90.0	&	472	\\
CPL	&$	345	\pm	53	$ &$	-0.16	\pm	0.14	$ &-- &$	4.3	\pm	1.2$ &-- &--&	188.6	/	201.1	/	 91.3	 &	 475	\\
CPL+BB	&$	359	\pm	63	$ &$	2.25	\pm	0.72	$ &-- &$	3.4	\pm	2.2$ &$	24	\pm	3	$ &$	0.8	\pm	0.3$&	189.9	 /	210.7	/	90.0	&	473	\\
$\textbf{(b) 0.394  -- 0.650}$&&&&&&\\	
BAND	&$	479	\pm	39	$ &$	-0.22	\pm	0.07	$ &$	-2.95	\pm	0.40	$ &$	19.4	\pm	3.8$ &-- &--&	 937.0	 /	953.7	/	464.5	&	474	\\
BAND+BB	&$	804	\pm	93	$ &$	-0.53	\pm	0.08	$ &$<-5$ &$	16.4	\pm	3.6$ &$	57	\pm	5	$ &$	 2.8	\pm	 1.1$&	 920.0	/	945.1	/	454.0	&	472	\\
CPL	&$	529	\pm	42	$ &$	-0.30	\pm	0.05	$ &-- &$	16.6	\pm	2.4$ &-- &--&	937.5	/	950.0	/	 465.8	&	 475	\\
CPL+BB	&$	804	\pm	136	$ &$	-0.53	\pm	0.08	$ &-- &$	15.6	\pm	4.0$ &$	57	\pm	5	$ &$	 2.8	\pm	 1.1$&	 918.0	/	938.9	/	454.0	&	473	\\
$\textbf{(c) 0.650  -- 0.782}$&&&&&&\\	
BAND	&$	331	\pm	24	$ &$	-0.02	\pm	0.10	$ &$	-2.96	\pm	0.40	$ &$	17.9	\pm	3.9$ &-- &--&	 -23.3	 /	-6.6	/	-15.6	&	474	\\
BAND+BB	&$	471	\pm	1	$ &$	-0.33	\pm	0.06	$ &$<-5$ &$	13.1	\pm	1.0$ &$	50	\pm	1	$ &$	 2.7	\pm	 0.6$&	 -22.7	/	2.3	/	-17.4	&	472	\\
CPL	&$	362	\pm	31	$ &$	-0.11	\pm	0.08	$ &-- &$	14.8	\pm	2.6$ &-- &--&	-22.4	/	-9.8	/	 -14.2	&	 475	\\
CPL+BB	&$	470	\pm	113	$ &$	-0.33	\pm	0.17	$ &-- &$	13.1	\pm	5.4$ &$	50	\pm	8	$ &$	 2.6	\pm	 1.6$&	 -24.7	/	-4.3	/	-17.4	&	473	\\
$\textbf{(d) 0.782 -- 1.138}$&&&&&&\\	
BAND	&$	200	\pm	18	$ &$	-0.20	\pm	0.11	$ &$	-3.04	\pm	0.61	$ &$	6.0	\pm	1.6$ &-- &--&	 1163.5	/	 1180.1	/	577.7	&	474	\\
BAND+BB	&$	245	\pm	20	$ &$	-0.17	\pm	0.20	$ &$	-4.03	\pm	1.31	$ &$	5.0	\pm	1.4$ &$	20	 \pm	 4	$ &$	0.4	\pm	0.3$&	1160.6	/	1185.6	/	574.3	&	472	\\
CPL	&$	214	\pm	16	$ &$	-0.26	\pm	0.07	$ &-- &$	4.9	\pm	0.7$ &-- &--&	1163.3	/	1175.8	/	 578.6	 &	 475	\\
CPL+BB	&$	246	\pm	40	$ &$	-0.17	\pm	0.20	$ &-- &$	4.6	\pm	1.6$ &$	19	\pm	4	$ &$	0.4	 \pm	0.3$&	 1159.3	/	1180.1	/	574.6	&	473	\\
$\textbf{(e) 1.138 -- 1.221}$&&&&&&\\	
BAND	&$	541	\pm	43	$ &$	-0.33	\pm	0.08	$ &$	-5.41	\pm	2.52	$ &$	19.8	\pm	3.1$ &-- &--&	 -593.2	/	-576.5	/	-300.6	&	474	\\
BAND+BB	&$	693	\pm	86	$ &$	-0.21	\pm	0.19	$ &$<-10$ &$	17.6	\pm	5.2$ &$	35	\pm	6	$ &$	 1.7	 \pm	 1.1$&	-602.4	/	-577.4	/	-307.2	&	472	\\
CPL	&$	542	\pm	67	$ &$	-0.33	\pm	0.08	$ &-- &$	18.7	\pm	3.7$ &-- &--&	-595.2	/	-582.6	/	 -300.6	&	 475	\\
CPL+BB	&$	693	\pm	141	$ &$	-0.21	\pm	0.20	$ &-- &$	17.8	\pm	6.9$ &$	35	\pm	6	$ &$	 1.7	 \pm	 1.2$&	-604.4	/	-583.6	/	-307.2	&	473	\\
$\textbf{(f) 1.221 -- 1.376}$&&&&&&\\	
BAND	&$	205	\pm	25	$ &$	-0.67	\pm	0.12	$ &$	-3.61	\pm	1.58	$ &$	3.4	\pm	0.9$ &-- &--&	 -136.6	/	 -119.9	/	-72.3	&	474	\\
BAND+BB	&$	253	\pm	67	$ &$	-0.78	\pm	0.23	$ &$<-10$ &$	2.9	\pm	1.5$ &$	22	\pm	9	$ &$	 0.1	\pm	 0.1$&	 -133.5	/	-108.5	/	-72.8	&	472	\\
CPL	&$	210	\pm	41	$ &$	-0.68	\pm	0.12	$ &-- &$	3.1	\pm	0.9$ &-- &--&	-138.4	/	-125.9	/	 -72.2	 &	 475	\\
CPL+BB	&$	253	\pm	112	$ &$	-0.79	\pm	0.24	$ &-- &$	3.0	\pm	1.5$ &$	22	\pm	9	$ &$	0.2	 \pm	0.1$&	 -135.5	/	-114.7	/	-72.8	&	473	\\
\hline
\end{tabular}
}
\end{center}
\end{table*}

\subsubsection{Time-Resolved spectral analysis}\label{time_resolved}
Time-resolved spectral analysis of GRB 170206A in six epochs is performed, such as [$T_0$+0.208 s, $T_0$+0.394 s] for epoch a, [$T_0$+0.394 s, $T_0$+0.650 s] for epoch b, [$T_0$+0.650 s, $T_0$+0.782 s] for epoch c, [$T_0$+0.782 s, $T_0$+1.138 s] for epoch d, [$T_0$+1.138 s, $T_0$+1.221 s] for epoch e and [$T_0$+1.221 s, $T_0$+1.376 s] for epoch f. In these spectral fittings, we set the initial spectral parameter values same as the resultant parameter values by spectral analysis in the GBM $T_{90}$ period.

Firstly, as seen in Table~\ref{tab:GRB170206A_spectra}, the time-resolved spectra in all epochs are not well fitted by the BAND+BB model due to the unconstrained high-energy photon index $\beta$ except for epoch d. Even in the epoch d, the BIC value derived by the model of BAND+BB is 9.9 larger than that in the models of CPL, which indicates a worse fit. Note that, BIC of BAND+BB model in each epoch is 6 larger than that of the model with the minimum BIC. Therefore, the BAND+BB model is rejected to fit the time-resolved gamma-ray spectra of GRB 170206A.

Secondly, we compare the BAND model and CPL model. With respect to the CPL model, the BAND model has $\Delta$BIC of 6.2, 6.1, and 6.0 in epochs a, e and f respectively, which implies that the CPL model is the better model. In other three epochs (b, c and d), the CPL model in each epoch has a smaller BIC value than that in the BAND model, however, the $\Delta$BIC is less than 6, such as 3.7, 3.2, and 4.3 respectively. With a minimum BIC in each epoch, thus we preferred the CPL model being a good model to fit all time-resolved spectra.

Finally, when comparing the CPL model and the CPL+BB model, the CPL+BB model is a better model to fit the spectrum than the CPL model in the epoch b with $\Delta$BIC = 11.1. The CPL model is a better model to fit the spectra in the epoch a ($\Delta$BIC = 9.6) and epoch f ($\Delta$BIC = 11.2). For epochs c and d, the CPL has smaller BIC values but with the $\Delta$BIC smaller than 6, such as 5.5 and 4.3 respectively, therefore we cannot reject the CPL+BB model in these two epochs. For epoch e, the CPL+BB model has a smaller BIC than that in the CPL model, i.e., $\Delta$BIC is 1.0, thus the CPL model is a compared model in this epoch.

In total, the CPL+BB model is the best-fit model in the epoch b, and could be a compared model in the epochs c, d and e. The CPL model is the best-fit model in epochs a and f, and could be a compared model in the epochs c, d and e.

\begin{figure}[p]
	\centering
	\subfigure[]{
		\includegraphics[width=0.45\textwidth]{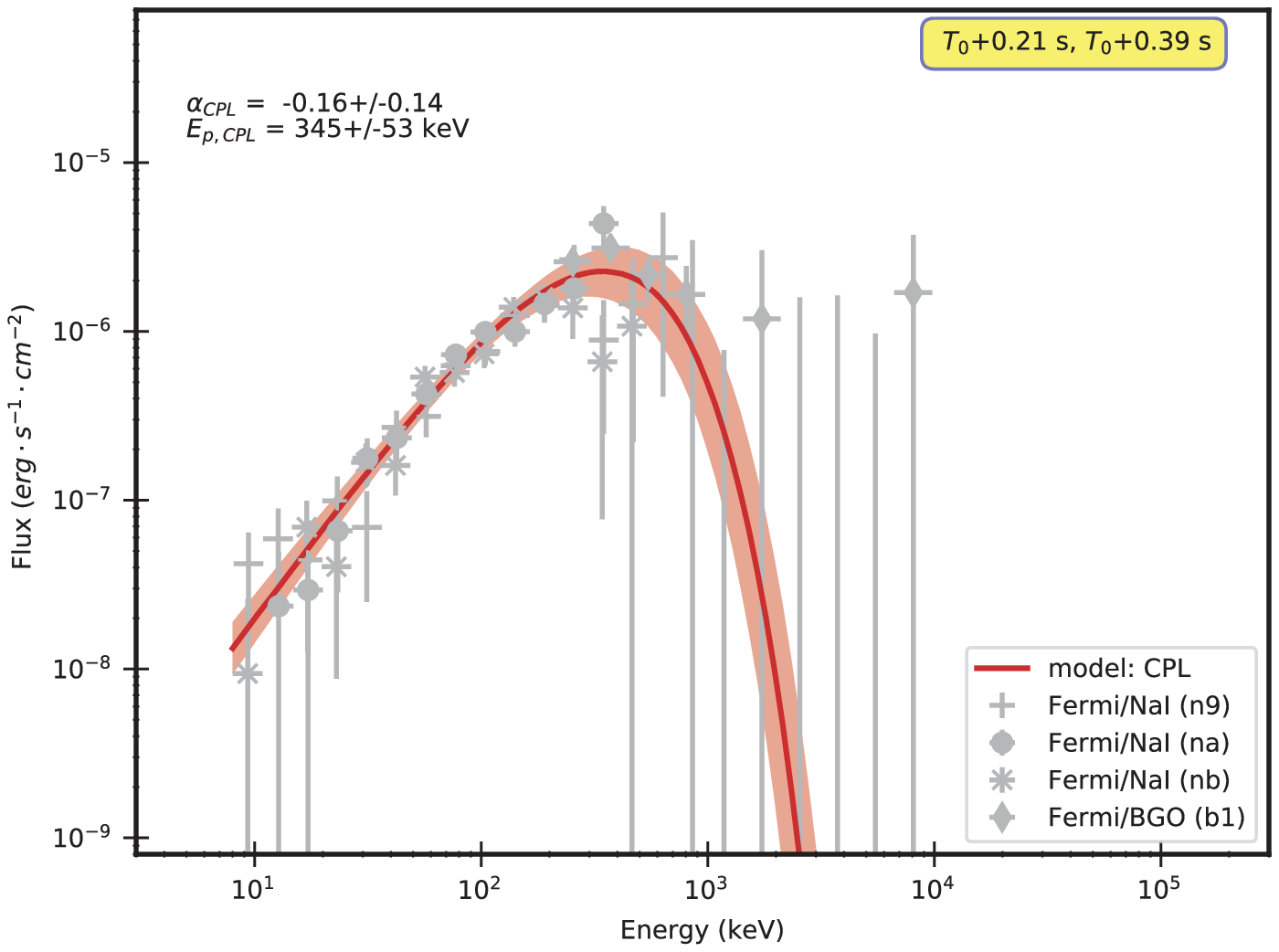}}
	\subfigure[]{
		\includegraphics[width=0.45\textwidth]{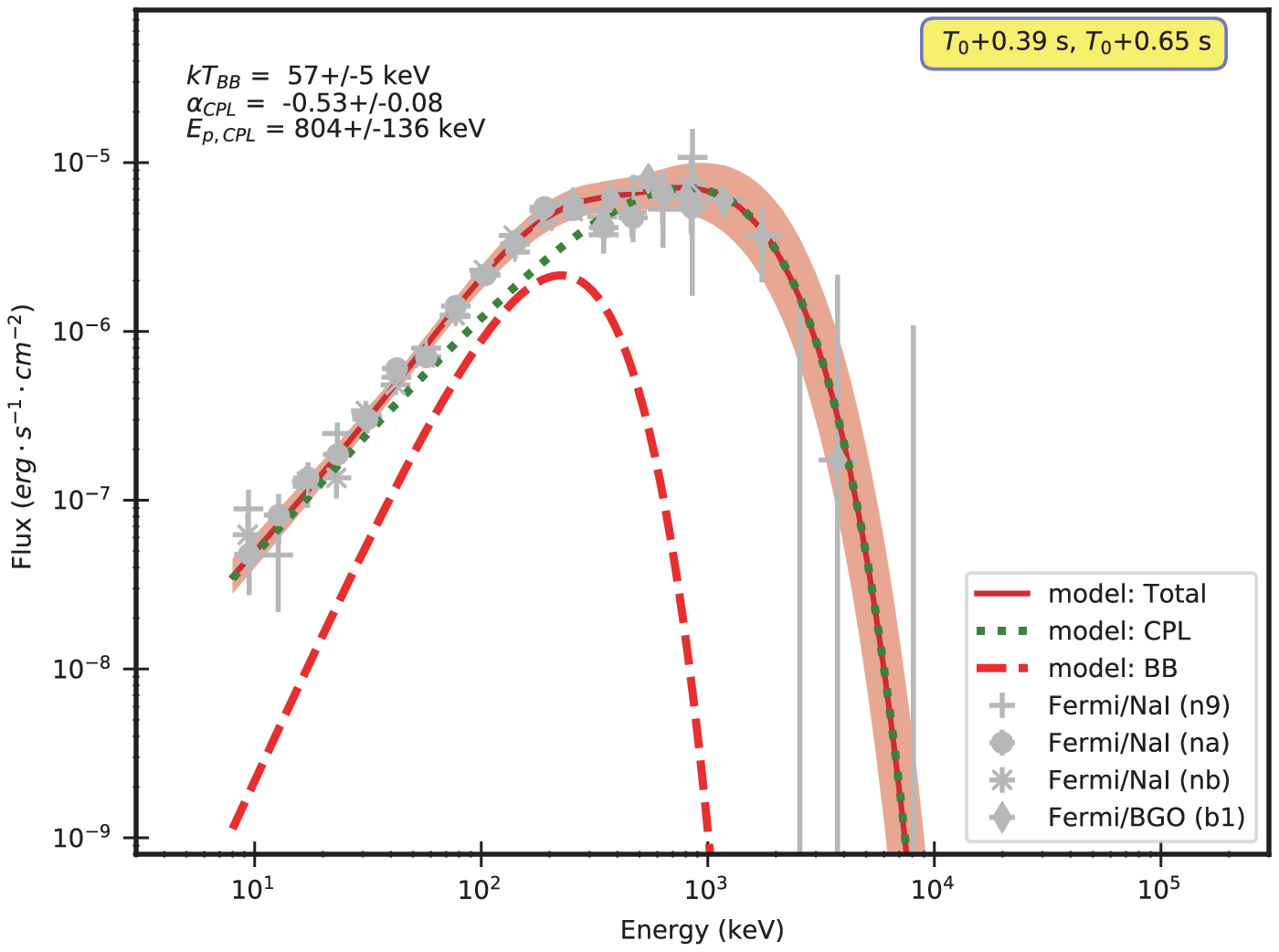}}
	\subfigure[]{
		\includegraphics[width=0.45\textwidth]{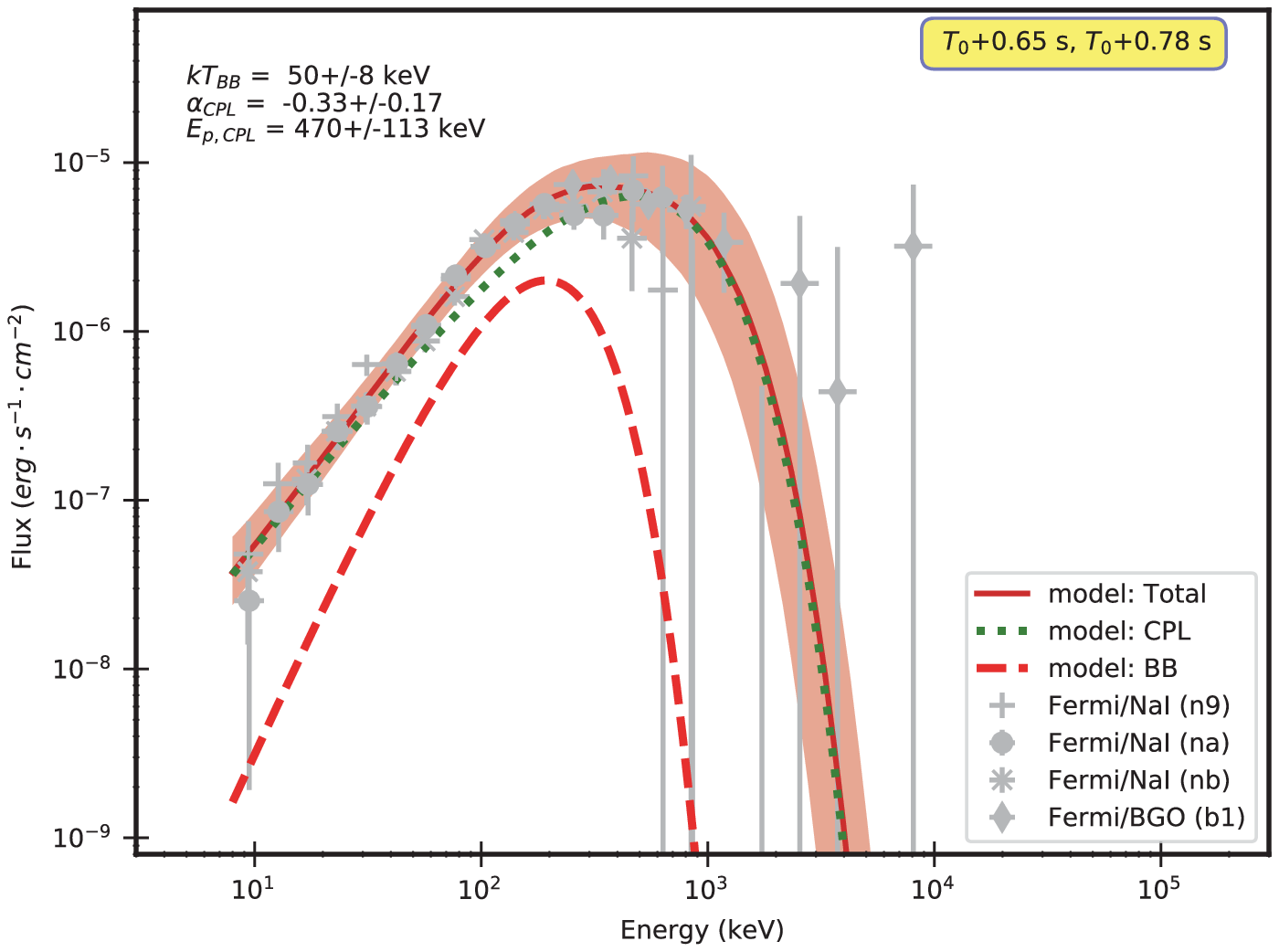}}
	\subfigure[]{
		\includegraphics[width=0.45\textwidth]{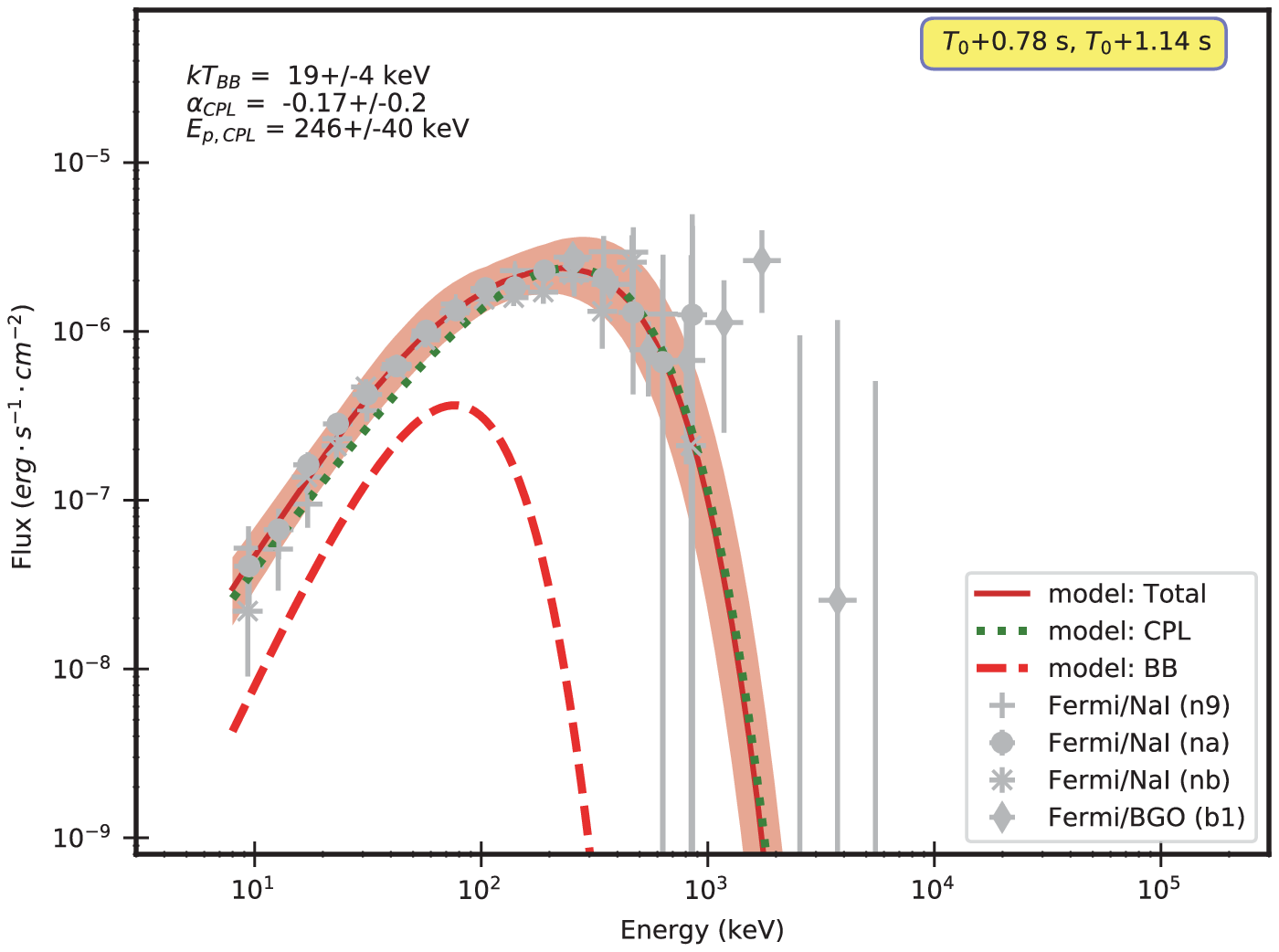}}
	\subfigure[]{
		\includegraphics[width=0.45\textwidth]{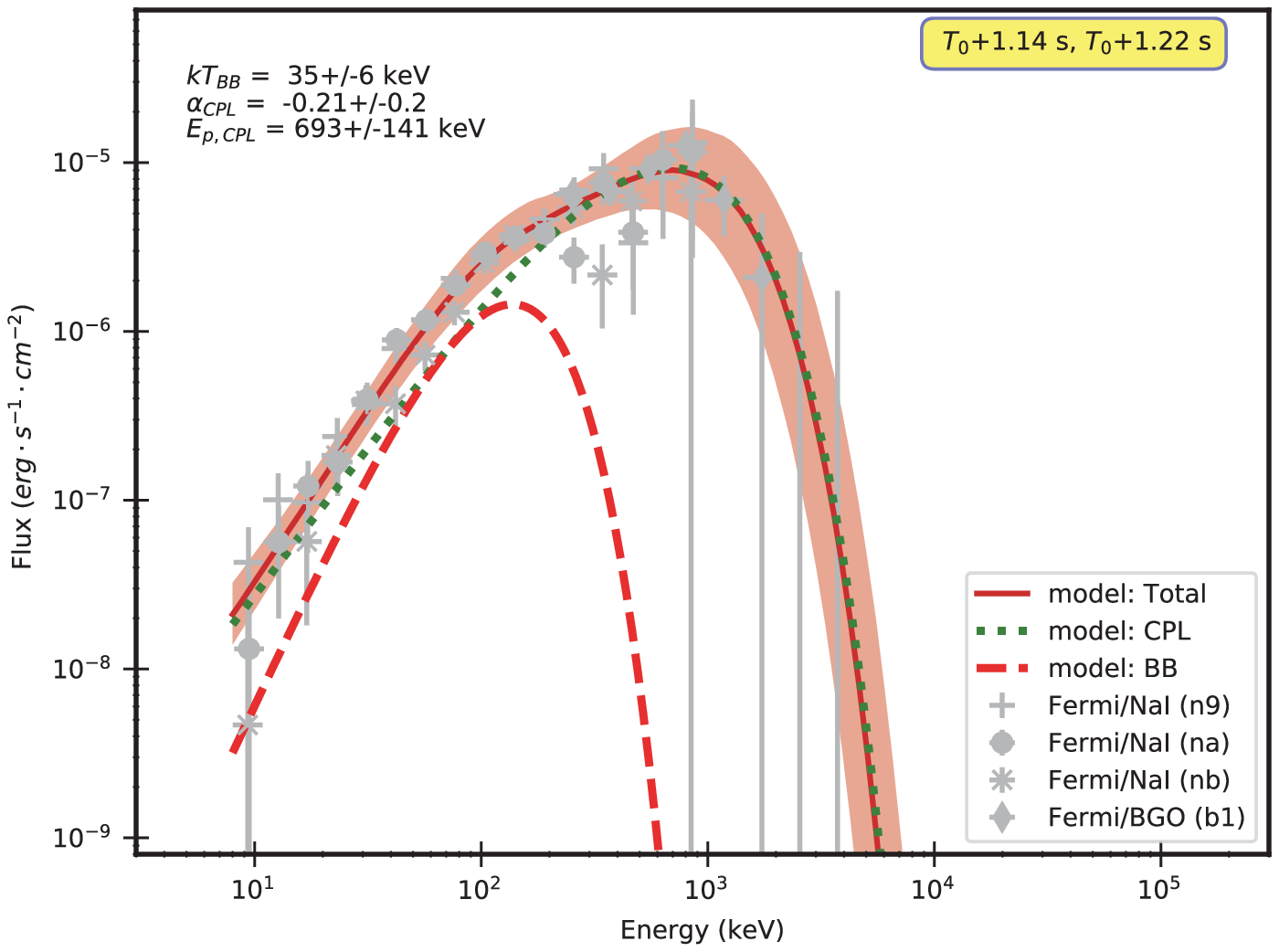}}
	\subfigure[]{
		\includegraphics[width=0.45\textwidth]{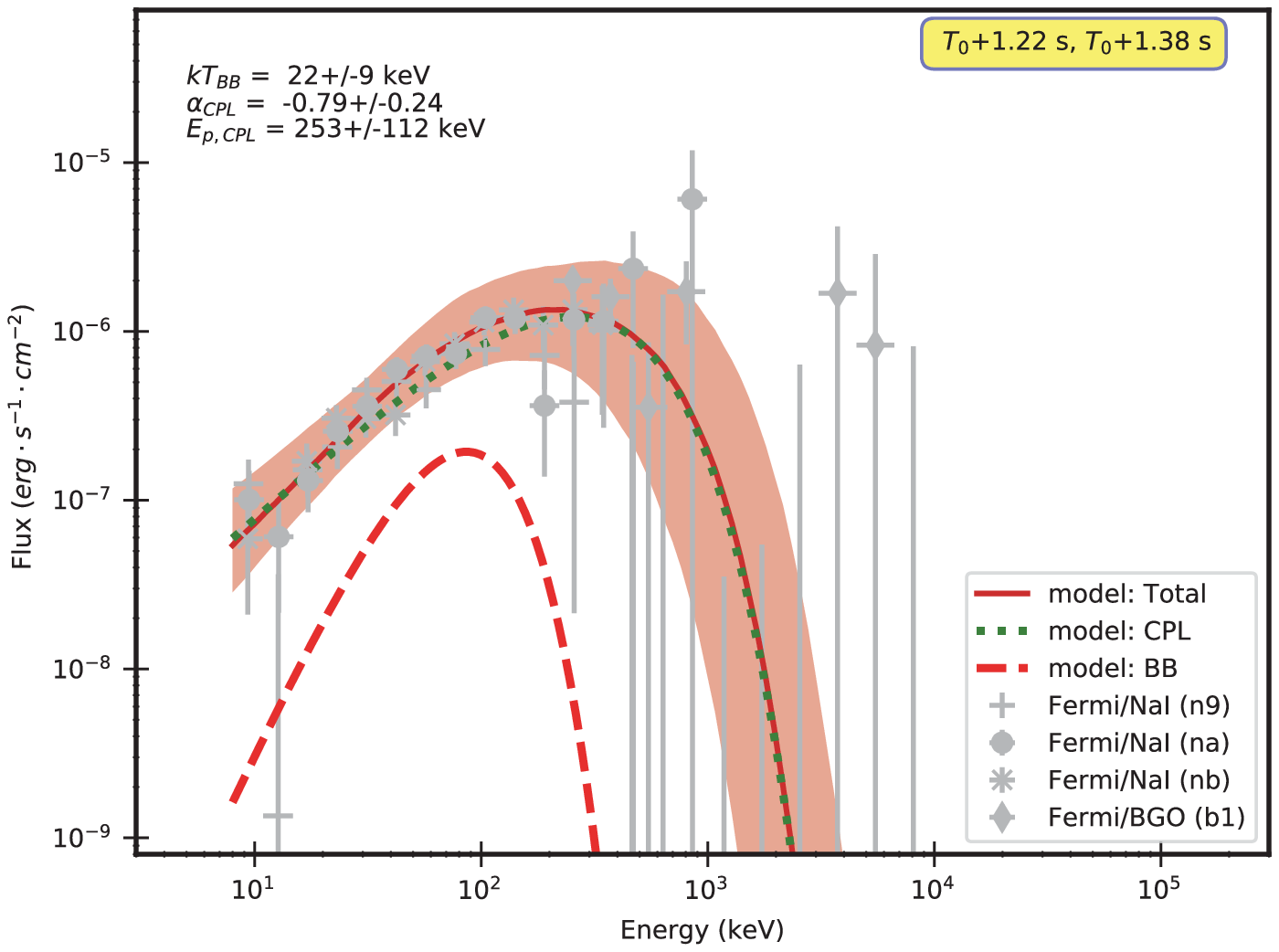}}
	\caption{\label{fig:sed2} Same as Figure 2, but for the time-resolved spectra of GRB 170206A during GBM $T_{90}$.}
\end{figure}

In order to discuss the parameter and flux variations during the GBM $T_{90}$, we, therefore, selecte the CPL+BB model as the fitting model in the following analysis excepted for epoch a, and all $\nu F\nu$ SEDs are plotted in Figure~\ref{fig:sed2}. Note that, for epoch a, the $\alpha_{\rm CPL}$ in the CPL+BB model is very hard, such as $+2.25(\pm0.72)$, thus finally we prefer the CPL model for epoch a.

\begin{figure*}
\centering
\includegraphics[width=0.70\textwidth]{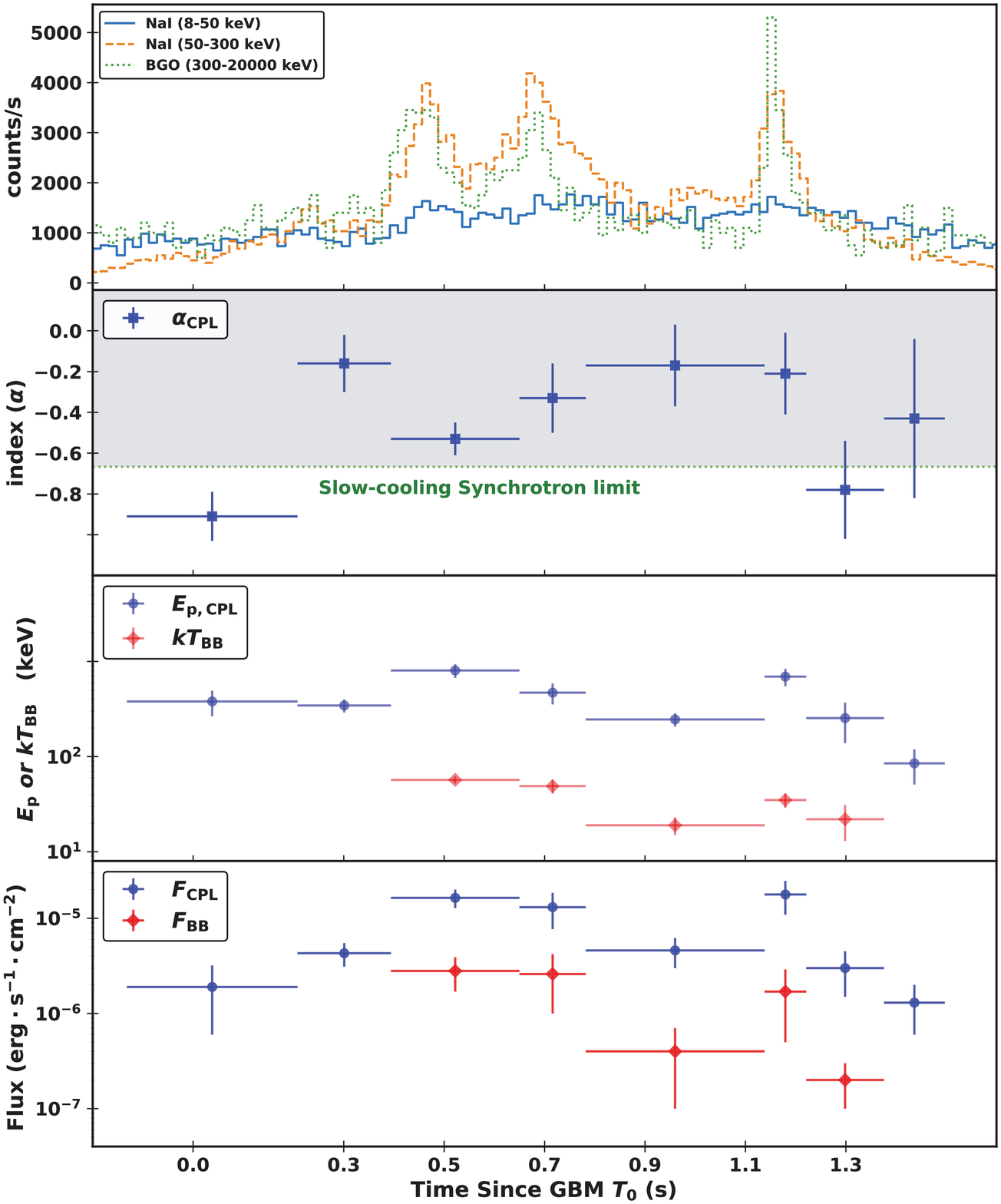}
\caption{\label{fig:para} Parameter value distributions as function of the time. Top panel: count-rates light curves for three energy bands. The second panel: the low-energy photon index, the gray shadow is the prohibit region of the synchrotron process. The third panel: the peak energy  ($E_{\rm p, CPL}$) in the $\nu F_\nu$ spectrum of the CPL component and the temperature of the BB component ($kT_{\rm BB}$). The bottom panel: the energy fluxes for the BB and CPL components.}
\end{figure*}

In Figure~\ref{fig:para}, temporal variations of the resultant parameters are plotted as well as the multi-wavelength GBM light curves. In the panel of the CPL index ($\alpha_{\rm CPL}$), the low-energy photon indices of epochs a, b, c, d and e are all out of the synchrotron limit (-2/3), which implies that the CPL component could not be of the standard synchrotron origin. For other epochs, such as Pre-$T_{90}$, f and Post-$T_{90}$, $\alpha_{\rm CPL}$ is also located more or less around the boundary of the synchrotron limit.

\begin{figure*}
\centering
\includegraphics[width=0.45\textwidth]{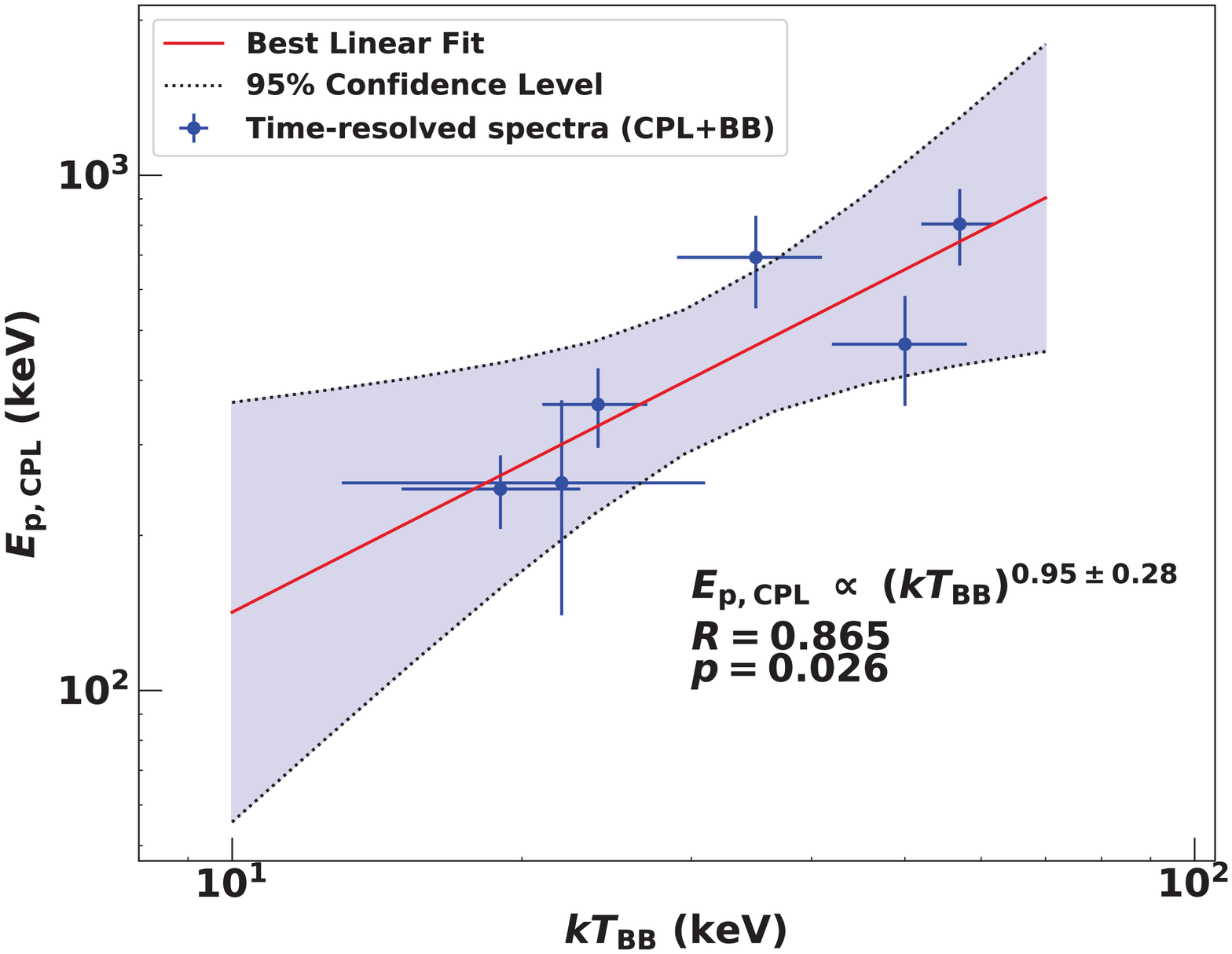}
\includegraphics[width=0.45\textwidth]{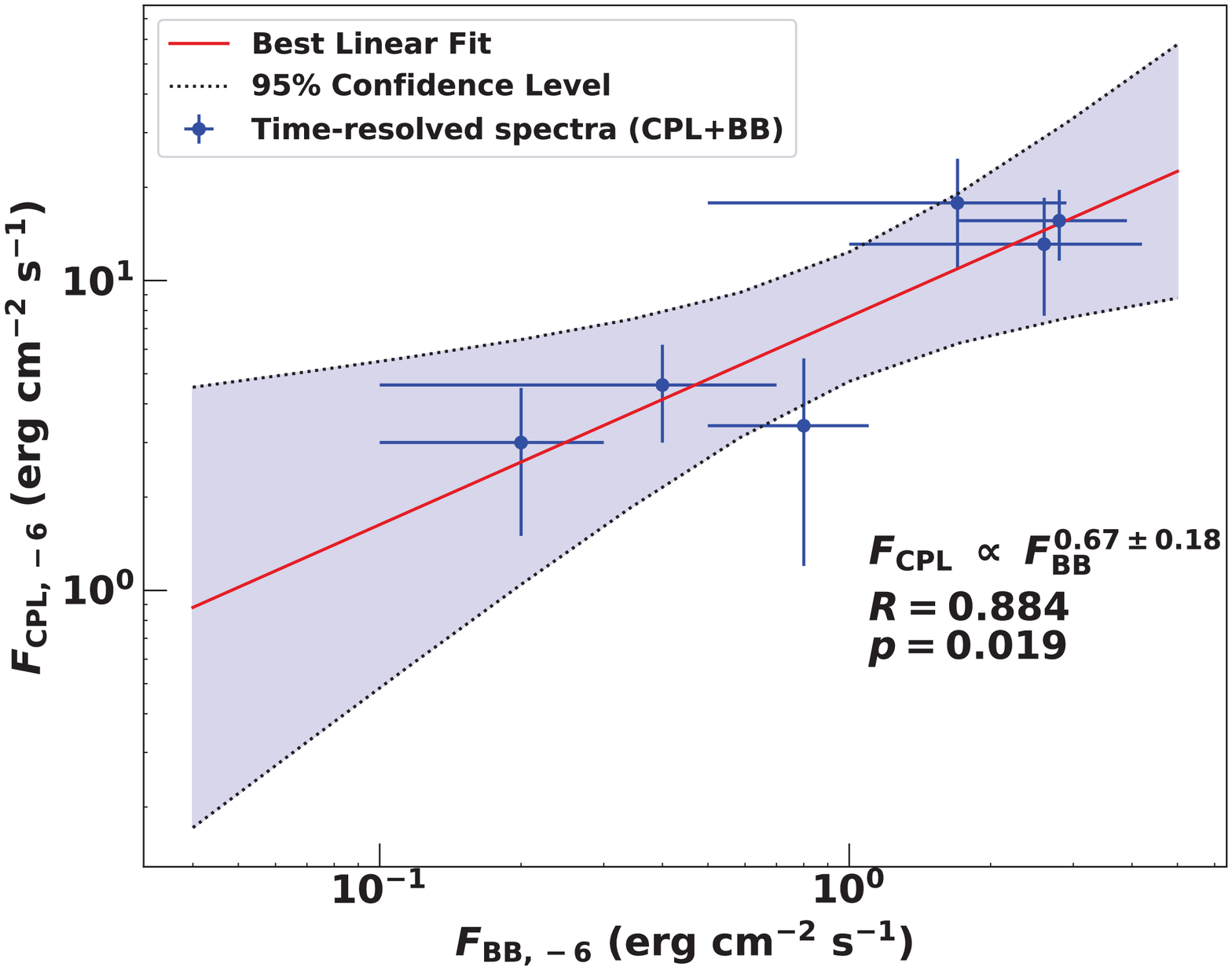}
\caption{\label{fig:correlation} Correlations between the time-resolved spectral parameters among GBM $T_{90}$ of GRB 170206A when fitted by the CPL+BB model. Left: $kT_{\rm BB}$ vs. $E_{\rm p, CPL}$. Right: $F_{\rm BB, -6}$ vs. $F_{\rm CPL, -6}$. The red lines are the best linear fitting and the shadows are the 95\% confidence levels of the best linear fitting. For each correlation, $R$ is the Pearson correlation coefficient and $p$ is the chance probability of the null hypothesis.}
\end{figure*}

For the peak energy of the CPL component ($E_{\rm p, CPL}$) and the temperature of the BB component ($kT_{\rm BB}$), they track each other well, such as decaying-rising-decaying. The correlation is tested in the time-resolved spectra employing the linear regression method in \textit{Origin} software package, which returns the Pearson correlation coefficient ($R$) and the chance probability of the null hypothesis ($p$). A strong positive correlation can be claimed when $R >0.8$ while a
moderate positive correlation can be claimed when $0.5< R <0.8$~\citep{Newton1999}. We find that $kT_{\rm BB}$ is strongly positively correlated with
$E_{\rm p, CPL}$, with $R = 0.865$ and $p = 0.026$ , as:
\begin{equation}
E_{\rm p,CPL} = 10^{1.20\pm0.42} kT_{\rm BB}^{0.95\pm0.28},
\label{epcpl}
\end{equation}
as seen in Figure ~\ref{fig:correlation}, where both $E_{\rm p,CPL}$ and $kT_{\rm BB}$ are in the unit of keV.

For the energy fluxes in time-resolved epochs derived from the CPL+BB model, the CPL fluxes ($F_{\rm CPL}$) also tracks the BB fluxes ($F_{\rm BB}$) well. The correlation analysis between them also favors a strong positive correlation, such as:
\begin{equation}
F_{\rm CPL,-6} = 10^{0.88\pm0.08} F_{\rm BB,-6}^{0.67\pm0.18},
\label{fcpl}
\end{equation}
with $R = 0.884$ and $p = 0.019$, which can be seen in Figure ~\ref{fig:correlation}. Here, $F_{-6} = 10^{-6}F$ and both fluxes are in the unit of ${\rm erg\ cm^{-2} \ s^{-1}}$.

\section{Origin of thermal and non-thermal components and its implications}

In addition to the four adopted spectral models in Table~\ref{tab:GRB170206A_spectra}, i.e., BAND, BAND+BB, CPL, and CPL+BB, we have compared other spectral models in Appendix~\ref{appendix} as well. As we can see, the other models do not present distinct advantages, so next, we focus on the four popular spectral models shown in Table~\ref{tab:GRB170206A_spectra} to explore the possible physical origins. Table~\ref{tab:GRB170206A_spectra} shows the fitting parameters of the spectra for the models of BAND, BAND+BB, CPL, and CPL+BB. When the BAND function is involved, either for the single BAND model or the BAND+BB model, usually, a very steep photon index at the higher energy band, namely, a very small $\beta$, has to be invoked. Such a small value of $\beta$ makes the BAND function approach the spectral shape of the CPL, implying that the real spectral shape may follow the CPL function rather than the typical BAND function. In addition, for the time-integrated and time-resolved spectra in most cases during the $T_{90}$ (see Section~\ref{specana}), one can see the CPL+BB model is fitting better compared with the single CPL model. Although in some cases, a single CPL model is good enough, this may be caused by the different weight of two components (BB and CPL components), inducing one component is overshot by another one. As a result, we take a more complicated observed spectral shape which contains two parts, i.e., a thermal component (the BB component) and a non-thermal component (the CPL component), to study their possible origins.

Besides, from the third and fourth panels of Figure \ref{fig:para}, one can see the plausible common evolution between the BB component and the CPL component, indicating a correlation between both components. Figure \ref{fig:correlation} shows their correlations, as seen in Equations \ref{epcpl} and \ref{fcpl}, which are stated as $E_{\rm p,CPL} \propto kT_{\rm BB}^{0.95\pm0.28}$, and $F_{\rm CPL} \propto F_{\rm BB}^{0.67\pm0.18}$.

Based on the above analyses, we suggest that the thermal emission and the non-thermal emission could imply two radiation regions \citep{Meszaros2002}. Basically, the thermal emission is a natural prediction from the photosphere of ``fireball'' model \citep{Meszaros2000,Meszaros2002,Rees2005}. Usually, the photons are coupled with the outflow due to the large optical depth at small radii and the spectrum emerging at the photosphere is shown as the blackbody distribution. Apart from this thermal emission from the photospheric origin, the non-thermal part could originate from the energy dissipation above the photosphere. Electrons above the photosphere could be accelerated to a non-thermal distribution. Thermal photons could serve as seed photons to Compton scattering of accelerated non-thermal energetic electrons above the photosphere and diverse setups of thermal photons could affect the final non-thermal spectrum emitted by these electrons \citep{Peer2005,Peer2006b,Peer2012,samuelsson2022}. In other words, the Comptonization of thermal photons shows as an additional non-thermal component to the thermal component. Such a connection between the thermal emission and the non-thermal emission may be responsible for the correlation between the BB component and the CPL component as shown in the third and fourth panels of Figure~\ref{fig:para}. Moreover, the low-energy spectral index of Comptonized photons, i.e., $\alpha$, could be harder than the death-line of synchrotron radiation (-2/3), inducing $\alpha$ ranging from -1.0 to 0.5 in some physical conditions \citep{Deng2014}. Such a range of $\alpha$ value is consistent with the low-energy photon indices listed in Table~\ref{tab:GRB170206A_spectra}, especially for those indices which are larger than $-2/3$ significantly.

Notice that the above suggested physical origin is based on the most preferred spectral functions, i.e., CPL or CPL+BB. The strong correlation between the BB component and the CPL component may be responsible for a single spectral function rather than two spectral functions, such as the mBB function mentioned in the Appendix although it has a worse BIC value. In this situation, the  suggested radiation model above will be invalid and the actual physical origin could be totally different \citep{ahlgren2015,vianello2018,samuelsson2022}.

\section{Conclusion}
In this work, we performed a comprehensive analysis of GRB 170206A with the observations by Fermi/GBM and Fermi/LAT in the prompt phase. A fast-variable thermal component is discovered, which has correlated photon fluxes with the non-thermal component throughout the $T_{90}$. Hard low-energy photon indices ($\alpha$) are found both in the time-integrated spectra and the time-resolved spectra. In the time-resolved spectra, the photon indices range from $-0.79$ to $-0.16$, most of which violate the line-of-death (-2/3) of the synchrotron slow-cooling radiation. In addition, we found the common evolution between the thermal component and the non-thermal component, indicating a positive correlation between photon fluxes as well as peak energies of both components. Based on the observational features, we explored the possible radiation models of GRB 170206A.

Assuming the two radiation regions for these two spectral components, the thermal component comes from the photosphere and the non-thermal component is from the Comptonization of the thermal component by the accelerated non-thermal energetic electrons above the photosphere. Since thermal photons serve as seed photons to Compton scattering of energetic electrons above the photosphere and thus affect the final non-thermal spectrum emitted by these electrons, the observational hard low-energy photon indices, as well as the positive correlation between their photon fluxes, can be reproduced.

\acknowledgments
We thank the anonymous referee for the constructive suggestions. We appreciate Francesco Capozzi and Ng Chun-Yu for improving the manuscript. This research made use of the High Energy Astrophysics Science Archive Research Center (HEASARC) Online Service at the NASA/Goddard Space Flight Center (GSFC). This work is supported by the NSFC under grants 11903017, 12065017, 11975116, and 12003007, the science research grants from the China
Manned Space Project with NO. CMS-CSST-2021-B11, Jiangxi Provincial Natural Science Foundation under grant 20212BAB201029, and the Fundamental Research Funds for the Central Universities (No. 2020kfyXJJS039).

\vspace{5mm}
\facilities{{\it Fermi}/GBM,{\it Fermi}/LAT}
\software{3ML\citep{2015arXiv150708343V}}

\appendix
\restartappendixnumbering

\section{Comparisons on ten spectral models in GRB 170206A \label{appendix}}

By including six more spectral models, there are ten spectral models are employed to fit the time-integrated and time-resolved SEDs of GRB 170206A and are selected to make comparisons. For example, the multicolor blackbody model (mBB), a single standard blackbody model (BB), double BB model (BB+BB), BB plus an additional powerlaw decay model (BB+PL), BAND with an additional PL model (BAND+PL) and CPL with an additional PL model (CPL+PL). For the mBB model, the same photon spectral function is employed as that in  \citet{Iyyani2021}, that is the model named \textit{diskpbb} in \textit{Xspec}, which can be written as,
\begin{equation}
    N(E)_{\rm mBB} = \frac{4 \pi E^2}{h^2 c^2} \left(\frac{A_{\rm mBB}}{\zeta}\right) T_{p}^{(2/\zeta)} \int_{T_{min}}^{T_{p}} \frac{T^{\frac{-(2+\zeta)}{\zeta}}}{e^{(E/T)} - 1} dT
\end{equation}
where $A_{\rm mBB}$ is the amplitude, $\zeta$ is power law index of the radial dependence of temperature ($ T(r) \propto r^{-\zeta}$), $T_{p}$ is the peak temperature in $keV$ and $T_{min}$ is the minimum temperature of the underlying blackbodies and is considered to be well below the energy range of the observed data, i.e., 8~keV in this work. For the PL function above, its photon model is presented as,
\begin{equation}
    N(E)_{\rm PL} = A_{\rm PL} \left(\frac{E}{100 \, \rm keV}\right)^{\Gamma}
\end{equation}
where $A_{\rm PL}$ is the amplitude, $\Gamma$ is the power law spectral index.

As seen in Table.~\ref{table:BICs}, there are three candidate models with $\Delta{\rm BIC}$ close to 0, that is mBB, CPL and CPL+BB. For the Pre-$T_{90}$ and Post-$T_{90}$ period, the CPL and the mBB models are the compared models. However, in the $T_{90}$ period, the CPL+BB model is the unique best model to fit its SED, which has none compared models. In the time-resolved spectra, the CPL model is the compared/best model in epochs a, c, d, e, and f, the mBB model is the compared/best model in epochs a, c, and f. The CPL+BB model is the compared/best model in epochs b, c, d and e.

\begin{figure*}
\centering
\includegraphics[width=0.75\textwidth]{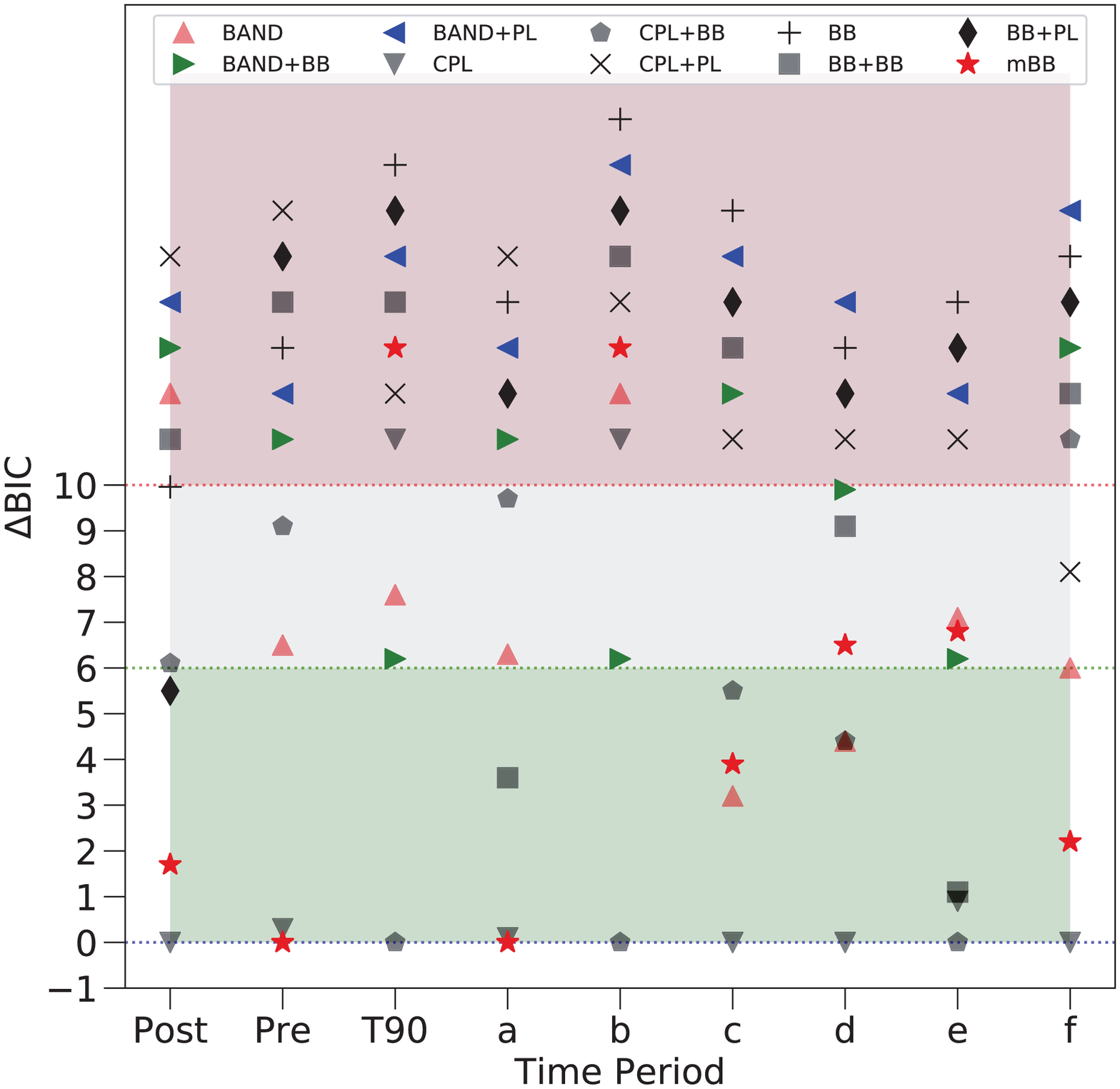}
\caption{\label{fig:BICs} $\Delta{\rm BIC}$ between each candidate model and the model with minimum BIC. Post, Pre and T90 represent the time-integrated spectra in Post-$T_{90}$, Pre-$T_{90}$ and $T_{90}$ period. The label a to e are the time-resolved spectra in six epochs between GBM $T_{90}$ duration. For three shadow regions, models in the dark red region (Top) with $\Delta{\rm BIC}$ larger than 10, thus all candidate models are rejected with a very strong evidence; models in the dark grey region (Middle, $6< \Delta{\rm BIC} <10$), the candidate models are not recommended with a strong evidence; models in the dark green region (Bottom, $\Delta{\rm BIC} < 6$) are the compared models with respect to the model with the minimum BIC.}
\end{figure*}

We did not present the result of the mBB model in the main text for two reasons. On the one hand, the mBB model is ruled out in $T_{90}$ period and three epochs (b, d and e), which includes two intensive main pulses, such as P1 and P3. On the other hand, the CPL model usually has a smaller BIC than that in the mBB model, such as in epochs c and f, even in epoch a, the mBB model has a BIC only 0.1 smaller than that in the CPL model. Therefore, we did not present the details of the mBB model in the main text. Although the BAND or BAND+BB model with $\Delta{\rm BIC}$ is mostly larger than 6 as seen in Figure~\ref{fig:BICs}, we include them in the main text since they are the popular models being considered in many published papers.

\begin{deluxetable*}{c|cccccccccc}
\tablecaption{$\Delta{\rm BIC}$ between each candidate model and the model with minimum BIC\tablenotemark{${\rm *}$}}
\label{table:BICs}
\tablehead{
\colhead{Period}
&\colhead{BAND}
&\colhead{BAND+BB}
&\colhead{BAND+PL}
&\colhead{CPL}
&\colhead{CPL+BB}
&\colhead{CPL+PL}
&\colhead{BB}
&\colhead{BB+BB}
&\colhead{BB+PL}
&\colhead{mBB}
}
\startdata
Post-T90	&	Unconstrained	&	Unconstrained	&	Unconstrained	&	0	&	6.1	&	Unconstrained	&	10.0	&	 22.3	&	5.5	&	1.7	\\
Pre-T90	&	6.5	&	33.1	&	34.2	&	0.3	&	9.1	&	Unconstrained	&	101.5	&	113.8	&	20.8	&	0	 \\
T90	&	7.6	&	6.2	&	339.7	&	14.4	&	0	&	26.7	&	1327.3	&	182.7	&	374.5	&	44.8	\\
\hline
a	&	6.3	&	15.9	&	39.6	&	0.1	&	9.7	&	Unconstrained	&	51.2	&	3.6	&	23.0	&	0	\\
b	&	14.8	&	6.2	&	143.0	&	11.1	&	0	&	23.5	&	385.5	&	63.6	&	138.5	&	22.0	\\
c	&	3.2	&	12.1	&	108.7	&	0	&	5.5	&	11.6	&	178.1	&	26.4	&	52.2	&	3.9	\\
d	&	4.4	&	9.9	&	Unconstrained	&	0	&	4.4	&	11.4	&	285.0	&	9.1	&	76.0	&	6.5	\\
e	&	7.1	&	6.2	&	68.9	&	0.9	&	0	&	13.3	&	183.6	&	1.1	&	101.2	&	6.8	\\
f	&	6.0	&	17.4	&	Unconstrained	&	0	&	11.2	&	8.1	&	112.6	&	14.1	&	20.7	&	2.2	\\
\enddata
	\tablenotetext{*}{$\Delta{\rm BIC}$ of the model with minimum BIC is presented as $0$.}
\end{deluxetable*}

\end{document}